\documentclass[prd,twocolumn,superscriptaddress,floatfix,nopacs,preprintnumbers,nofootinbib]{revtex4}
\usepackage[utf8]{inputenc}

\usepackage{graphicx}
\usepackage[normalem]{ulem}
\usepackage{xcolor}
\usepackage{adjustbox}
\usepackage{mathrsfs}
\usepackage{amssymb,bm}
\usepackage{amsmath}
\usepackage{mathtools}
\usepackage{physics}
\usepackage{slashed}
\usepackage{verbatim}
\usepackage{ragged2e}
\usepackage{caption}
\usepackage{subcaption}
\usepackage{multirow}
\usepackage{array}
\newcommand{\nc}{N_\mathrm{c}}
\newcommand{\aem}{\alpha_\mathrm{em}}
\newcommand{\gev}{\mathrm{GeV}}

\newcommand{\bt}{\mathbf{b}}
\newcommand{\qso}{Q_{\mathrm{s},0}} 
\newcommand{\qs}{Q_{\mathrm{s}}} 
\newcommand{\rt}{\mathbf{r}}

\newcommand{\kt}{\mathbf{k}}

\newcommand{\xt}{\mathbf{x}}

\newcommand{\xbj}{x_\mathrm{Bj}}

\newcommand{\lqcd}{\Lambda_\text{QCD}}

\newcommand{\xij}[1]{\mathbf{x}_{#1}}

\newcommand{\btheta}{\boldsymbol{\theta}}
\newcommand{\ymodel}{\mathbf{y}(\btheta)}
\newcommand{\ymodelc}{\text{y}(\btheta)}
\newcommand{\yexp}{\mathbf{y}_{\mathrm{exp}}}
\newcommand{\deltay}{\Delta \mathbf{y}(\btheta)}

\usepackage[breaklinks,colorlinks,citecolor=citcolor,urlcolor=blue,linkcolor=lcolor]{hyperref}
\definecolor{lcolor}{rgb}{0.5,0,0}
\definecolor{citcolor}{rgb}{0,0.3,0.0}
\definecolor{teal}{rgb}{0.0, 0.5, 0.5}
\newcommand{\as}{\alpha_\mathrm{s}}
\newcommand{\der}{\mathrm{d}}

\begin{document}

\title{Inferring the initial condition for the Balitsky-Kovchegov equation}

\author{Carlisle Casuga}
\email{carlisle.doc.casuga@jyu.fi}
\author{Mikko Karhunen}
\email{mikko.a.karhunen@student.jyu.fi}
\author{Heikki Mäntysaari}
\email{heikki.mantysaari@jyu.fi}
\affiliation{
Department of Physics, University of Jyväskylä,  P.O. Box 35, 40014 University of Jyväskylä, Finland
}
\affiliation{
Helsinki Institute of Physics, P.O. Box 64, 00014 University of Helsinki, Finland
}

\begin{abstract}
    We apply Bayesian inference to determine the posterior likelihood distribution for the parameters describing the initial condition of the small-$x$ Balitsky-Kovchegov evolution equation at leading logarithmic accuracy. The HERA structure function data is found to constrain most of the model parameters well. In particular, we find that the HERA data prefers an anomalous dimension $\gamma\approx 1$ unlike in previous fits where $\gamma>1$ which led to e.g. the unintegrated gluon distribution and the quark-target cross sections not being positive definite. The determined posterior distribution can be used to propagate the uncertainties in the non-perturbative initial condition when calculating any other observable in the Color Glass Condensate framework. We demonstrate this explicitly for the inclusive quark production cross section in proton-proton collisions and by calculating predictions for the nuclear modification factor for the $F_2$ structure function in the EIC and LHeC/FCC-he kinematics.
\end{abstract}

\maketitle

\section{Introduction}

Understanding the high-energy structure of protons and nuclei is one of the central goals of the next-generation Electron-Ion Collider (EIC)~\cite{AbdulKhalek:2021gbh}. Especially the fact that one can perform nuclear-DIS experiments for the first time in collider kinematics is intriguing: at high center-of-mass energies it is possible to access the high-density part of the nuclear wave function where non-linear QCD dynamics with emergent saturation phenomena is expected to play a major role.

Precise theory calculations taking into account saturation effects are necessary to probe in detail the non-linear QCD dynamics in hadronic collisions at RHIC and at the LHC, as well as in accurate nuclear-DIS experiments~\cite{Morreale:2021pnn} at the EIC or at the proposed future LHeC/FCC-he collider~\cite{LHeC:2020van}.
The Color Glass Condensate (CGC) effective theory of QCD~\cite{Gelis:2010nm} provides a natural framework to describe gluon saturation  at high energies. In recent years there has been a rapid progress towards next-to-leading order accuracy in CGC calculations, see e.g. Refs.~\cite{Boussarie:2016bkq,Beuf:2021srj,Beuf:2022ndu,Shi:2021hwx,Chirilli:2012jd,Stasto:2013cha,Caucal:2023fsf,Bergabo:2023wed,Ducloue:2017ftk,Beuf:2017bpd,Mantysaari:2022kdm}. In addition to higher order corrections, precise theoretical predictions require a well constrained non-perturbative input describing the proton or nuclear structure at moderately small-$x$, which is an initial condition to the perturbative evolution equations such as the Balitsky-Kovchegov (BK)~\cite{Kovchegov:1999yj,Balitsky:1995ub} equation resumming contributions enhanced by  large logarithms of center-of-mass energy.

The initial condition for the BK evolution has been fitted to the HERA structure function data~\cite{H1:2009pze,H1:2015ubc} at leading~\cite{Albacete:2009fh,Lappi:2013zma,Albacete:2010sy,Ducloue:2019jmy} and recently also at next-to-leading~\cite{Hanninen:2022gje,Beuf:2020dxl} order accuracy (see also Ref.~\cite{Dumitru:2023sjd} for a complementary approach to determine the initial condition from the large-$x$ structure of the proton). This initial condition is a necessary input when calculating any other scattering process at small-$x$, such as particle production in proton-nucleus collision at the LHC, see e.g. Refs.~\cite{Albacete:2010bs,Lappi:2013zma,Shi:2021hwx,Mantysaari:2023vfh}.

All current fits to the DIS data extracting the initial condition for the BK evolution lack an uncertainty analysis, i.e.  do not provide any method to propagate the uncertainties in the initial condition parametrization to the calculated observables. When developing the CGC calculations to the precision level, a statistically rigorous treatment of these uncertainties can be seen to be of equal importance as the higher order corrections when comparing predictions  to current and future experimental data. 

In this work we extract, for the first time with an uncertainty estimate, the non-perturbative initial condition for the BK evolution from the 
HERA structure function data~\cite{H1:2015ubc}.
This is achieved by employing a Bayesian inference setup extensively used in the field, recently when extracting e.g. properties of the Quark Gluon Plasma~\cite{Bernhard:2016tnd,Bernhard:2019bmu,Nijs:2022rme,Soeder:2023vdn,Parkkila:2021yha,JETSCAPE:2020shq} and the event-by-event fluctuating proton geometry~\cite{Mantysaari:2022ffw}. The determined posterior distribution for the initial condition  parameters allows for a rigorous propagation of initial condition parametrization uncertainties to calculations of all other observables in the CGC framework. In this work we present results based on a leading order analysis but note that the developed computationally efficient setup can be extended to the numerically heavy next-to-leading order case, where it has been recently shown that combined analyses including both total cross section and charm production data are feasible~\cite{Hanninen:2022gje}. 

This paper is structured as follows. In Sec.~\ref{sec:dis} we review the calculation of the proton structure functions at high energy in the CGC approach. The Bayesian inference setup is described in Sec.~\ref{sec:bayesian}. The determined initial condition for the BK evolution with an uncertainty estimate is presented in Sec.~\ref{sec:results}. To illustrate the propagation of uncertainties to  calculations of various observables, we calculate in Sec.~\ref{sec:applications} inclusive quark production in proton-proton collisions and the nuclear modification factor for the structure function $F_2$ which will be measured at the EIC but for which no predictions that include the BK evolution consistently with the HERA data exist in the literature (see however Ref.~\cite{Armesto:2022mxy}). Conclusions are presented in Sec.~\ref{sec:conclusions}, and the importance of including correlations among the experimental uncertainties is analyzed in Appendix~\ref{appendix:correlations}.

\section{Deep inelastic scattering at high energy}
\label{sec:dis}

The total cross section in lepton-nucleus scattering is typically expressed in terms of the reduced cross section defined as 
\begin{multline}
    \sigma_r(\xbj,y,Q^2) = F_2(\xbj,Q^2)
    \\
    - \frac{y^2}{1 + (1-y)^2} F_L(\xbj,Q^2).
\end{multline}
Here $y=Q^2/(sx)$ is the inelasticity, $\xbj$ is the Bjorken-$x$, $Q^2$ the photon virtuality and $\sqrt{s}$ the electron-proton center-of-mass energy. The structure functions $F_2$ and $F_L$ are related to the total virtual photon-nucleus cross section as
\begin{align}
    F_2 &= \frac{Q^2}{4\pi^2 \aem} (\sigma^{\gamma^* A}_T + \sigma^{\gamma^* A}_L) \\
    F_L &= \frac{Q^2}{4\pi^2 \aem} \sigma^{\gamma^* A}_L.
\label{eq:f2fl}
\end{align}
Here the subscripts $T$ and $L$ refer to the transverse and longitudinal virtual photon polarization states, respectively.

In the dipole picture the total cross section for the virtual photon-nucleus scattering at high energy factorizes to a product of the virtual photon wave function $\psi^{\gamma^*\to q\bar q}$ describing $\gamma^*\to q\bar q$ fluctuation, and the dipole-target scattering amplitude $N$ as~\cite{Kovchegov:2012mbw}
\begin{multline}
\label{eq:lo-cross-section}
    \sigma^{\gamma^* p}_{T,L}=2 \sum_f \int \der^2 \bt \der^2 \rt \frac{\dd z}{4\pi} |\psi^{\gamma^* \to q\bar q}(\rt,Q^2,z)|^2
    \\
    \times
    N(\rt,\bt,\xbj).
\end{multline}
Here $z$ is the fraction of the photon longitudinal momentum carried by the quark, $\rt$ is the transverse size of the $q\bar q$ dipole (higher Fock states would enter beyond leading order) and $\bt$ is the center-of-mass of the dipole. We sum over light quarks $f=u,d,s$, as it has been shown in Ref.~\cite{Albacete:2010sy} that in a leading order calculation with the energy dependence obtained from the BK evolution
it is not possible to simultaneously describe the total and charm production cross section data.

Following Refs.~\cite{Lappi:2013zma,Albacete:2010sy,Albacete:2009fh}, when calculating the virtual photon-proton cross section we neglect the impact parameter dependence in the dipole-proton amplitude $N$ and replace
\begin{equation}
    \int \dd[2]\bt \to \frac{\sigma_0}{2}.
\end{equation}
Here the proton transverse area $\sigma_0/2$ is taken to be a free parameter. 

The squared photon wave functions summed over quark helicities read~\cite{Kovchegov:2012mbw} 
\begin{multline}
\label{eq:photon_t}
	|\psi_T(r,z,Q^2)|^2 = \frac{2\nc}{\pi} \aem e_q^2 \Big\{ [ z^2+(1-z)^2] \varepsilon^2 K_1^2(\varepsilon r) \\
	+ m_f^2 K_0^2(\varepsilon r) \Big\}
\end{multline}
and
\begin{equation}
\label{eq:photon_l}
	|\psi_L(r,z,Q^2)|^2 = \frac{8 \nc}{\pi} \aem e_q^2 Q^2 z^2(1-z)^2 K_0^2(\varepsilon r),
\end{equation}
with $\varepsilon^2 = z(1-z)Q^2 + m_q^2$. Here $m_q=0.14$~GeV is the light quark mass and $e_q$ is the quark charge. 
One could also in principle take the light quark mass to be a fit parameter as e.g. in Refs.~\cite{Albacete:2010sy, Mantysaari:2018nng}. However these previous analyses have found only a small sensitivity on $m_q$ and as such we choose a fixed value that is compatible with  previous fits.

The Bjorken-$x$ dependence of the dipole amplitude is given by the Balitsky-Kovchegov equation:
\begin{multline}
    \label{eq:bk-evolution}
    \frac{\partial N(\xij{01})}{\partial Y} =  \int \der^2 \xij{2} K_\text{BK}
    (\xij{0}, \xij{1}, \xij{2})
    \\
    \times
    [N(\xij{02}) + N(\xij{12}) - N(\xij{01}) - N(\xij{02})N(\xij{12})].
\end{multline}
Here $\xt_{ij}=\xt_i-\xt_j$ and $Y=\ln x_0/x$ (see also Ref.~\cite{Ducloue:2019ezk} for a detailed discussion of the evolution variable). In this work we initiate the BK evolution at $x_0=0.01$. When the running coupling corrections are included following the Balitsky prescription~\cite{Balitsky:2006wa} the kernel reads 
\begin{multline}
\label{eq:bk-rc-balitsky}
  K_\text{BK}(\xij{0},\xij{1}, \xij{2}) = \frac{\nc \as(\xij{01}^2)}{2\pi^2} \left[
        \frac{\xij{01}^2}{\xij{12}^2 \xij{02}^2} \right. \\
        + \frac{1}{\xij{02}^2} \left( \frac{\as(\xij{02}^2)}{\as(\xij{12}^2)} -1 \right) 
        + \left. \frac{1}{\xij{12}^2} \left( \frac{\as(\xij{12}^2)}{\as(\xij{02}^2)} -1 \right)
  \right].
\end{multline}
In this work we present a leading-order analysis. 
We note that in addition to next-to-leading order corrections~\cite{Beuf:2017bpd,Ducloue:2017ftk,Beuf:2021srj,Beuf:2022ndu}, it is also possible to resum corrections enhanced by large transverse logarithms to the BK equation following Refs.~\cite{Iancu:2015vea,Iancu:2015joa,Beuf:2014uia} (that in practice approximate the full NLO BK equation accurately~\cite{Balitsky:2008zza,Lappi:2016fmu}). Furthermore, it would be worthwhile to use the recently developed resummation procedure from Ref.~\cite{Ducloue:2019ezk} where the evolution is formulated in terms of the target rapidity instead of the projectile rapidity. These improvements we leave for future work.

For the initial condition of the BK evolution we use a McLerran-Venugopalan model~\cite{McLerran:1993ni} inspired parametrization as e.g. in Ref.~\cite{Lappi:2013zma}
\begin{multline}
    \label{eq:bk-ic}
    N(\xij{ij}, x=x_0) =
    1-\exp 
        \left[  - \frac{\left(\xij{ij}^2 \qso^2\right)^\gamma}{4} \right. \\
            \times \left. \ln \left( \frac{1}{|\xij{ij}| \Lambda_\text{QCD}}
             +  e_c\cdot  e   \right)   \right]. 
\end{multline}
Here the free parameters are $\qso^2$ related to the initial saturation scale, the anomalous dimension $\gamma$ controlling the shape of the dipole amplitude at small $|\xij{ij}|$ and the infrared regulator $e_c$. The  Bayesian inference setup discussed in Sec.~\ref{sec:bayesian} can efficiently handle multidimensional parameter space, which allows us to use this more flexible parametrization for the initial condition compared to functional forms used in previous fits (e.g. in Ref.~\cite{Lappi:2013zma} either $\gamma$ or $e_c$ was fixed).

The strong coupling constant in Eq.~\eqref{eq:bk-rc-balitsky} depends on the transverse distance scale. Following again Refs.~\cite{Albacete:2009fh,Lappi:2013zma} we write the coordinate space strong coupling as
\begin{equation}
    \label{eq:bk-ic-rc}
    \as(\rt^2) = \frac{12\pi}{(33-2n_f) \ln\left( \frac{4C^2}{\rt^2 \lqcd^2} \right)}.
\end{equation}
Here we take $n_f=3$ as we only consider light quarks in this work, and $\lqcd=0.241\,\gev$. In the infrared region the coupling is frozen to $\as=0.7$.  The parameter $C^2$ controls the strong coupling scale in the coordinate space, and based on Refs.~\cite{Balitsky:2006wa,Kovchegov:2006vj} its expected value is $C^2=e^{-2\gamma_E}\approx 0.32$. However we consider $C^2$ to be a free parameter to control the uncertainty of the strong coupling scale and to absorb some missing higher order corrections, for example the fact that the next-to-leading order evolution is known to result in slower evolution speed compared to the leading order case~\cite{Lappi:2016fmu}. Previous leading order fits have also been found to prefer a much larger $C^2= 7\dots 15$~\cite{Lappi:2013zma}.

\section{Bayesian analysis setup}
\label{sec:bayesian}

The non-perturbative parameters describing the initial condition for the BK evolution, the proton transverse area and the scale of the coordinate space running coupling can be estimated using Bayesian parameter inference. In this work we use a similar inference procedure as e.g. in Refs.~\cite{Mantysaari:2022ffw,Bernhard:2016tnd,Parkkila:2021yha}.
The output is a multidimensional likelihood distribution that comprehensively describes the parameter uncertainties and correlations based on prior beliefs and experimental data. 

Bayesian inference is based on Bayes' theorem:
\begin{equation}
    P(\ymodel | \yexp) \propto P(\yexp | \ymodel) P(\btheta).
    \label{eq:bayesian}
\end{equation}
Here the model output vector with given parametrization $\boldsymbol{\theta} = (\qso^{2}, \gamma, e_{c}, C^{2}, \sigma_0/2) $ over $N$ kinematical points $\mathbf{x}_i = (\xbj, Q^{2}, y)_i$ is represented by $\ymodel= (\ymodelc_{\mathbf{x}_1}, \ymodelc_{\mathbf{x}_2}, \dots ,\ymodelc_{\mathbf{x}_N}$). The posterior function, represented by $P(\ymodel|\yexp)$, is the probability of $\ymodel$ being true given the experimental data, $\yexp$, as evidence. 

The probability represented by $P(\yexp|\ymodel)$  is a measure of the agreement between the observation and the model calculation at the kinematical point $\mathbf{x}$ given the model parameters $\boldsymbol{\theta}$. 
We use a multivariate normal distribution in which case the logarithm of the likelihood function reads
\begin{multline}
    \label{eq:loglikelihood}
    \log P(\mathbf{y}_\mathrm{exp} | \mathbf{y}(\boldsymbol{\theta})) 
    \\ 
    = -\frac{1}{2} \left[ \deltay^{T} \Sigma^{-1}(\boldsymbol{\theta}) \deltay + \log ( 2\pi \mathrm{det} \hspace{1mm} \Sigma) \right].
\end{multline}
Here $\Sigma = \Sigma_\mathrm{model}(\btheta) + \Sigma_\mathrm{exp}$ is a matrix sum of the model, or emulator as discussed shortly, and the experimental covariance matrices and the difference $\deltay = \mathbf{y}(\btheta) - \mathbf{y}_{\mathrm{exp}}$ is a vector with length $N$. The experimental data considered in this work is the reduced cross section data measured by the H1 and ZEUS collaborations~\cite{H1:2015ubc}, where the published dataset also includes the full covariance matrix describing correlations among the 162 sources of systematic uncertainty. 
Furthermore we include as correlated uncertainties the 7 different procedural uncertainties originating from the combination of the H1 and ZEUS datasets. 
These correlations between different systematic uncertainties have not been included in  previous dipole model fits, and the effect of these correlations on the final posterior distribution is illustrated in Appendix~\ref{appendix:correlations}.
The uncorrelated systematic uncertainties and statistical uncertainties are added in quadrature.

The prior, represented by $P(\btheta)$ in Eq.~\eqref{eq:bayesian}, encodes initial knowledge on the range and shape of the probability distribution of the model parameters. Not much information is known about the true distribution of the model parameters other than estimates of previous fits from Refs.~\cite{Lappi:2013zma, Albacete:2010sy,Albacete:2009fh}. In this study, the prior is chosen to be a flat multivariate distribution with bounds corresponding to guesses of possible parameter ranges, outside of which the posterior is set to zero. Initially, the width of the posterior distribution is unknown, so the study conservatively begins with a wide allowable extent and is later narrowed down. The  prior ranges used in the final analysis are shown in Table~\ref{tab:medianmap}.

\begin{figure}[tb]
    \centering
    \includegraphics[width=0.45\columnwidth]{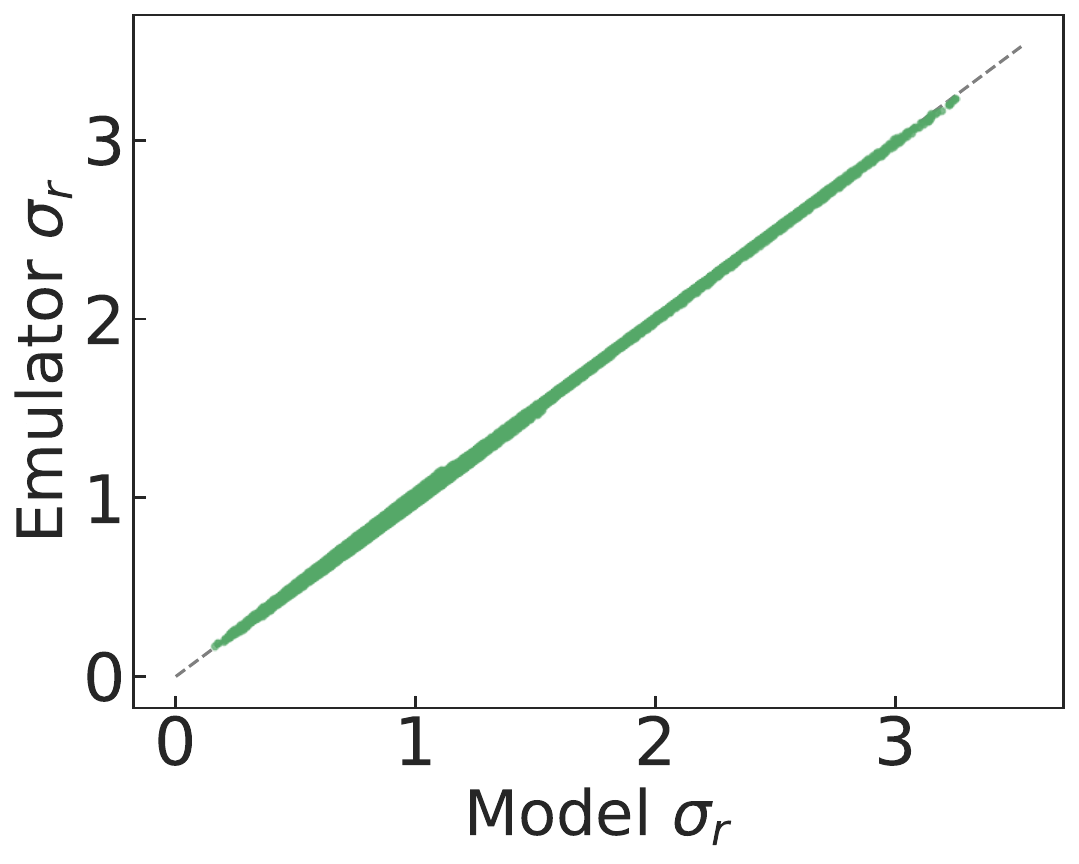}
    \includegraphics[width = 0.43\columnwidth]{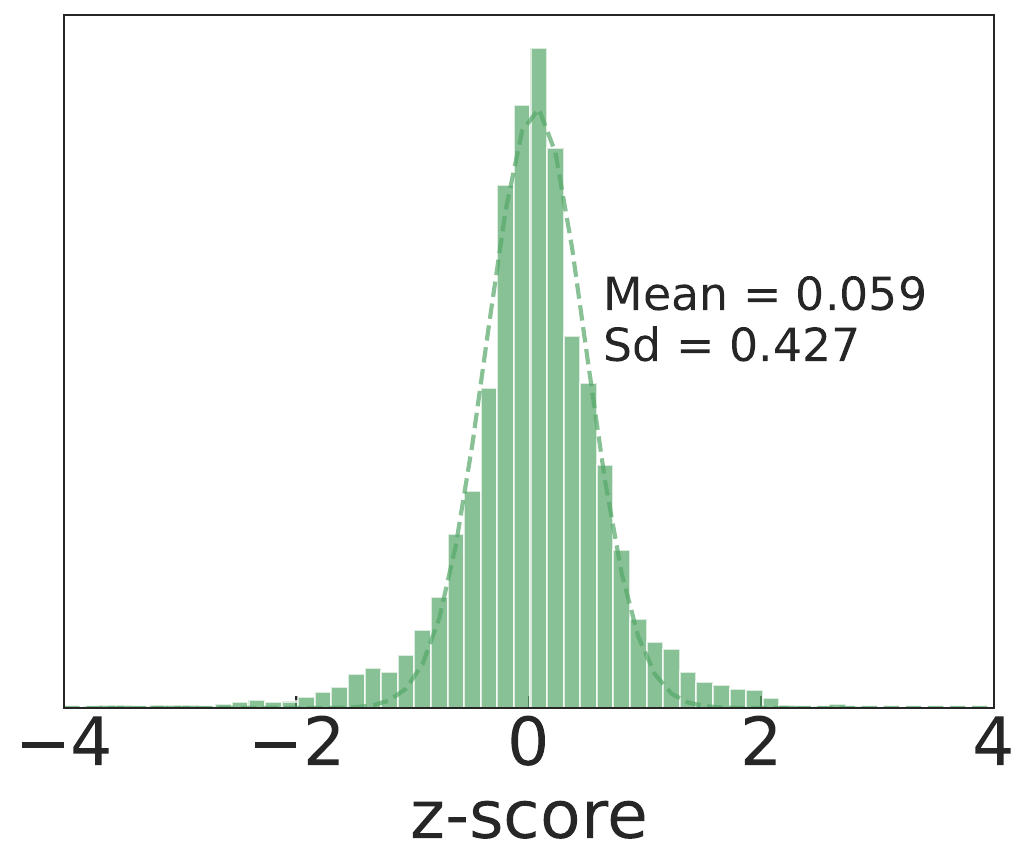}
    \caption{Validation for the 5-parameter Bayesian fit (left) comparing the calculated reduced cross section to the emulator prediction. The $z$-score plot (right) is constructed with 500 validation points over 403 kinematics. The dashed line shows a Gaussian fit to the $z$-score histogram, for which the mean and the standard deviation (SD) are given in the figure.  }
    \label{fig:validation}
\end{figure}

In order to effectively sample the multidimensional parameter space, we employ  Gaussian Process Emulators (GPEs)~\cite{scikit-learn} 
as computationally efficient surrogates for the actual theory calculation, and the Markov Chain Monte Carlo (MCMC) sampling method. 
The MCMC importance sampling is implemented via the \texttt{emcee} package~\cite{emcee} that uses an affine-invariant ensemble sampler.  This  sampler uses an algorithm similar to the Metropolis-Hastings (MH)  where each step of a walker from $A \rightarrow B$ is accepted with a probability of $q \sim \frac{P(\theta_{B})}{P(\theta_{A})}$. Unlike the MH algorithm, \texttt{emcee} employs an ensemble of $k$ \textit{walkers} where the advancement of the chain of one walker is also based on the current positions of the other walkers; therefore, it converges much faster than MH. Each step is then appended to the MCMC chain to, eventually, form the target posterior distribution. 

The GPEs are trained at different points in the parameter space sampled from a latin hypercube which ensures that the design points are optimally distributed in the parameter space. This significantly reduces the number of training points needed to get accurate estimates from GPEs, and in this analysis we use 500 training parametrizations.
We include data in the region $2.0\,\mathrm{GeV}^{2}< Q^{2} < 50.0\,\mathrm{GeV}^{2}$ where there are 403 datapoints. The low-$Q^2$ cut ensures the presence of a perturbative scale, and the upper limit removes contribution from the region dominated by DGLAP dynamics not included in the present analysis. We represent the HERA data in this kinematical domain by 6 principal components that capture 99.99999\% variance of the model output within the prior.
The GP output at a certain $(\mathbf{x}, \btheta)$ is a mean estimate and a covariance matrix that describes the emulator uncertainties and is used to form the matrix $\Sigma_\mathrm{model}$.

The GPEs are validated by comparing their predictions with the actual model calculations of the reduced cross section. The left panel of Fig.~\ref{fig:validation} show the comparison between the model and the emulator output at all kinematical $(\xbj,Q^2,y)$ points in the HERA data computed using separate validation sets of model parameters not included in the training of the GPEs. The emulator accuracy is found to very good, with the average relative difference being $0.047\%$ in our standard setup where $\qso^2, e_c, C^2, \sigma_0/2$ and $\gamma$ are free parameters. The right-hand side panel of Fig.~\ref{fig:validation} shows the corresponding $z$-score defined as
\begin{equation}
    z = \frac{\mathrm{GPE} - \mathrm{model}}{\sigma_\mathrm{GPE}},
\end{equation}
where $\sigma_\mathrm{GPE}$ is the uncertainty estimate of the GPE.
The width of the $z$ distribution is less than unity which means that the emulator uncertainties are typically slightly overestimated. However, it is the experimental uncertainty that dominates in the likelihood function.

\section{Inferred initial condition for the BK evolution}

\label{sec:results}
\begin{figure*}[ht]
    \centering
    \includegraphics[width=\textwidth]{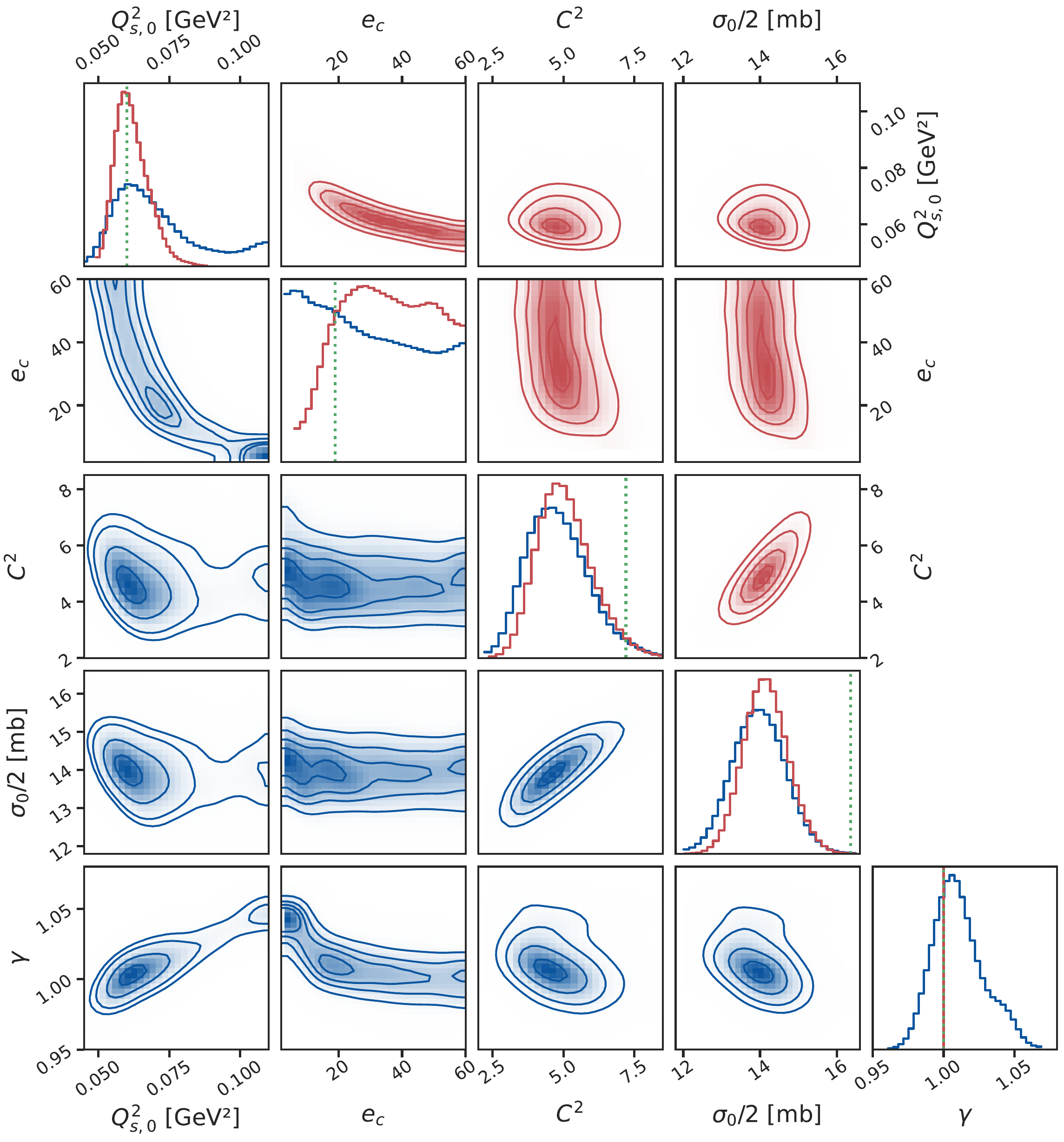}
    \caption{1- and 2-dimensional projections of the posterior probability distributions for both the 4-parameter (red) and 5-parameter case (blue). The dotted green line shows the best fit values from the MV\textsuperscript{e} fit of Ref.~\cite{Lappi:2013zma}}. 
    \label{fig:corner}
\end{figure*}

\renewcommand{\arraystretch}{1.5} 
\begin{table*}[ht]
    \centering
    \begin{tabular}{|p{5cm}|wc{2cm}|wc{2cm}|wc{2cm}|wc{2cm}|wc{2cm}|}
    \hline
        \multirow{2}{*}{Parameter Description} & \multirow{2}{*}{Prior Range} & \multicolumn{2}{c|}{4-parameter case} & \multicolumn{2}{c|}{5-parameter case}  \\ \cline{3-6}
          & & Median & MAP & Median & MAP \\
    \hline
    \hline
        Initial scale, $\qso^2$ [GeV$^{2}$] & [0.04, 0.11] &  $0.061_{-0.007}^{+0.014}$ & 0.060 &  $0.067_{-0.015}^{+0.041}$ & 0.077 \\
        
        Infrared regulator, $e_c$ & [0.5, 60.0] & $35.3_{-23.7}^{+23.3}$ & 38.9 & $27.5_{-23.0}^{+31.0}$ & 15.6 \\
        
        Running coupling scale, $C^{2}$ & [2.0, 10.0] & $4.97_{-1.40}^{+2.22}$ & 4.60 & $4.72_{-1.58}^{+2.52}$ & 4.47 \\
        
        Proton transverse area, $\sigma_{0}/2$ [mb] & [12.0, 18.0] & $14.1_{-1.0}^{+1.1}$ & 13.9 & $14.0_{-1.2}^{+1.3}$ & 13.9 \\
        
        Anomalous dimension, $\gamma$ & [0.9, 1.1] & 1 (fixed) & 1 (fixed) & $1.01_{-0.03}^{+0.04}$ & 1.01 \\
    \hline
    \multicolumn{2}{|c|}{Saturation scale, $Q_s^{2}$, at $\xbj=0.01$ $[\mathrm{GeV}^2]$} & 0.279 & 0.280 & 0.288 & 0.289 \\ 
        \multicolumn{2}{|c|}{$\chi^{2}$/d.o.f.} & 1.013 & 1.011 & 1.016 & 1.012\\
    \hline
    \end{tabular}
    \caption{MAP and median values for the 4- and 5-parameter setups. The uncertainty estimates in the 95\% credible intervals are also shown.} 
    \label{tab:medianmap}
\end{table*}

We determine the posterior distribution of model parameters in two separate setups. First, in our main setup we consider all parameters discussed above ($Q_{\mathrm{s},0}^2,e_c,C^2,\sigma_0/2$ and $\gamma$) to be free and determine the corresponding posterior distribution. We will refer to this setup as the 5-parameter case. For comparison we also consider the case where we set $\gamma=1$ as in the standard MV model, and refer to this as the 4-parameter setup. This parametrization corresponds to the ``MV\textsuperscript{e}'' one used in Ref.~\cite{Lappi:2013zma}. 

The obtained posterior distributions are shown in Fig.~\ref{fig:corner} for both the 4- and 5-parameter cases. 
The plots along the diagonal present the 1D projections of the posterior showing the likelihood distributions for individual parameters. The off-diagonal plots are 2D histograms that illustrate the correlations between the pairs of model parameters. 

\begin{figure}
    \centering
    \includegraphics[width=\columnwidth]{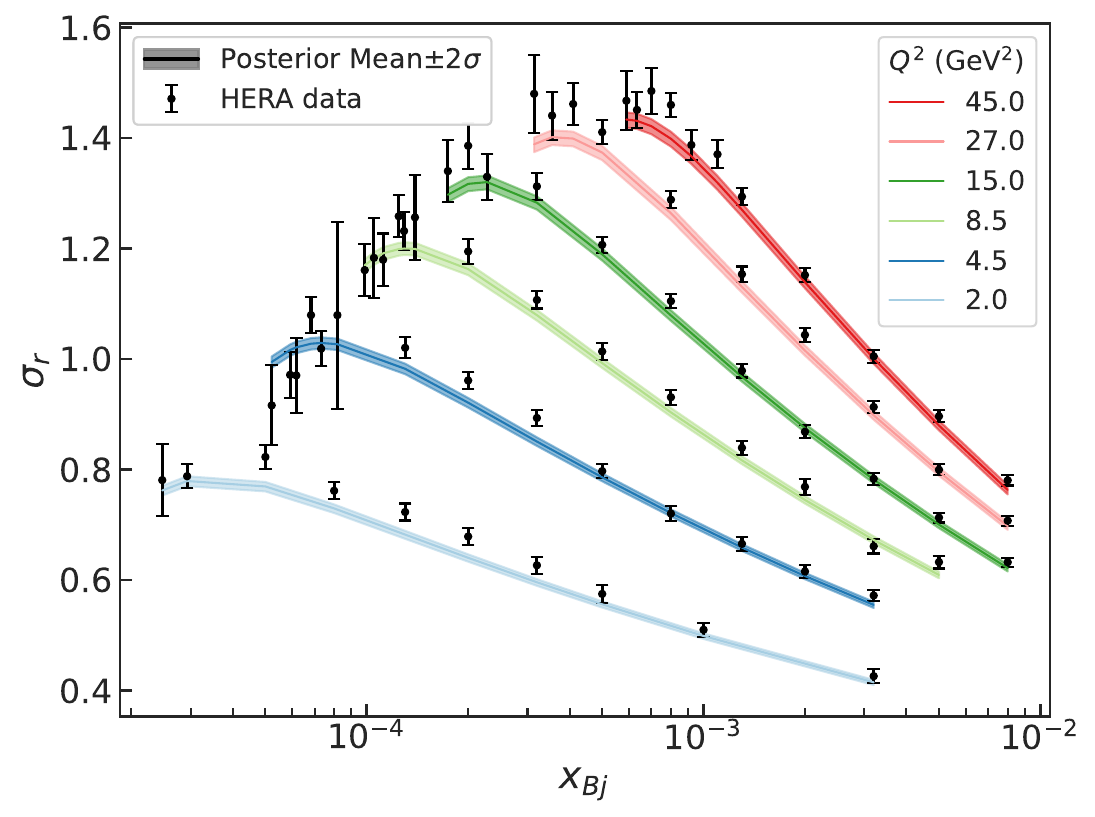}
    \caption{The reduced cross section as a function of $\xbj$ and the $2\sigma$ uncertainty band obtained from the posterior samples in selected $Q^2$ bins compared with the HERA data at $\sqrt{s}=318\,\mathrm{GeV}$~\cite{H1:2015ubc}.} 
    \label{fig:reducedcrosssection}
\end{figure}

The HERA data is found to constrain most of the model parameters well, except the infrared regulator $e_c$ for which the likelihood distribution extends to large values of $e_c$ in both setups. The broad distribution is not unexpected as the dipole amplitude depends on $e_c$ weakly and only in the region where $r \gtrsim 1/\Lambda_\mathrm{QCD}$, see Eq.~\eqref{eq:bk-ic}. Also, in the 5-parameter case, large values for $Q_{\mathrm{s},0}^2$ are allowed and compensated by small $e_c$, and in both setups we obtain a clear negative correlation between $Q_{\mathrm{s},0}^2$ and $e_c$.

In the 5-parameter case where $\gamma$ is a free parameter, we find that the HERA data prefers $\gamma=0.95 \dots 1.05$. This is in contrast to previous fits performed in Refs.~\cite{Lappi:2013zma, Albacete:2010sy,Albacete:2009fh} where the $Q^2$-dependence of the HERA data was shown to prefer $\gamma\approx 1.1\dots 1.2$ (in fits with fixed $e_c=1$). 
As we will discuss in more detail in Sec.~\ref{sec:applications}, having $\gamma\lesssim 1$ is advantageous e.g. when calculating inclusive particle production cross section in proton-nucleus collisions.
The most likely values for the individual parameters in the 4-parameter case are similar to what is reported in the previous leading order fit~\cite{Lappi:2013zma}.  

In addition to the negative $Q_{\mathrm{s},0}^2$-$e_c$ correlation discussed above, we also find a clear negative correlation between $Q_{\mathrm{s},0}^2$ and $\sigma_0/2$ especially in the 5-parameter case. This is because, in the region where the virtual photon-proton cross section is dominated by small dipoles, one has  
\begin{equation}
    \sigma^{\gamma^* p} \sim \frac{\sigma_0}{2} N(r) \sim \frac{\sigma_0}{2} \left(\frac{Q_{\mathrm{s},0}^{2}}{Q^2}\right)^{\gamma},  
\end{equation}
where we use the fact that $r^2\sim 1/Q^2$ (except in the aligned jet limit where $z\approx 0$ or $z\approx 1$). This dependence also explains the negative correlation seen between $\gamma$ and $\sigma_0/2$. 

\begin{figure}[h!]
    \centering
    \includegraphics[width=\columnwidth]{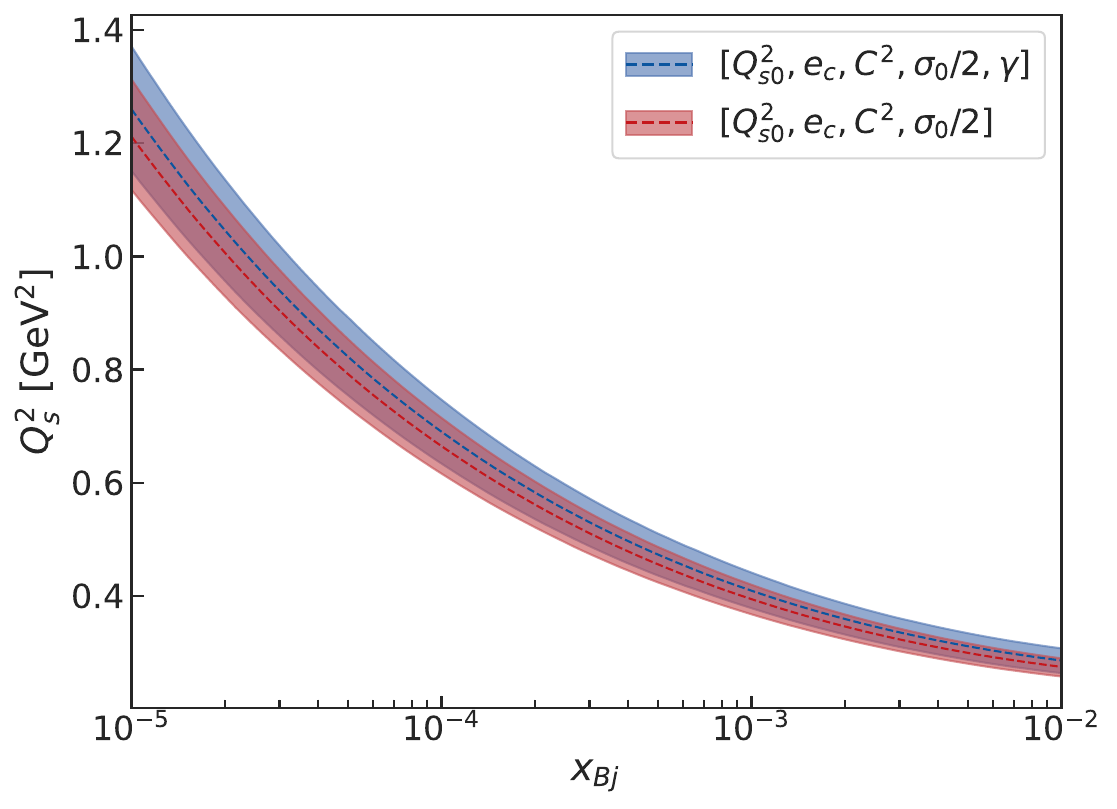}
    \caption{The saturation scale as a function of $\xbj$, defined as the solution to $N(\rt) = 1 - e^{-1/2}$ when $ \rt^2 =2/Q_s^{2}$. This is shown for both the 4- and 5- parameter cases with $2\sigma$ uncertainty bands.} 
    \label{fig:qs2ofx}
\end{figure}

As stated earlier, $\qso$ parametrizes  the saturation scale at initial $\xbj$. The saturation scale, $\qs$, determines the scale at which non-linear dynamics becomes important. Independent of the functional form of the BK initial condition, we can define (following e.g.~\cite{Lappi:2013zma}) $\qs^2$ as the solution to $N(\rt^{2} = 2/\qs^{2}, \xbj) = 1 - e^{-1/2}$. The determined saturation scales at initial $\xbj$ in both 4- and 5-parameter setups are shown in Table.~\ref{tab:medianmap}, and are found to be identical in both cases. Furthermore, Fig.~\ref{fig:qs2ofx} shows the $\xbj$ dependence of $\qs^2$, which is identical in both setups despite the differences in the posterior distributions.

At large $\qso^{2}$, the typical running coupling $\as(r) \sim \as(1/\qs)$ is smaller. Consequently in that case there is no need for as large $C^{2}$ to get a slow enough evolution speed compatible with the $x$-dependence of the HERA data. Hence the negative correlation between $C^{2}$ and $\qso^2$ is obtained, and consequently a positive correlation between $C^2$ and $\sigma_0/2$. These correlations appear very clearly in both  setups. Consequently, it is essential to take into account the correlations between the model parameters when estimating uncertainties when calculating different cross sections. 

The median values for the parameters are presented in Table~\ref{tab:medianmap} along with the \textit{maximum a posteriori} (MAP) estimates corresponding to the maximum of the posterior distribution. 

The compatibility with the HERA data is quantified by $\chi^2/\mathrm{d.o.f}$ defined as
\begin{equation}
    \frac{\chi^2}{\mathrm{d.o.f}} = \frac{1}{N-p} \deltay^{T} \Sigma_\mathrm{exp}^{-1} \deltay, 
\end{equation}
where $N=403$ is the number of experimental datapoints in the considered kinematical domain and $p=4$ or $p=5$ is the number of free model parameters. The $\chi^2/\mathrm{d.o.f}\approx 1.01$ shows that an excellent description of the precise HERA data is obtained, similarly as in previous leading order fits. By calculating the average cross section over many samples from the posterior and not using a single parametrization (e.g. median or MAP), we get $\chi^2/\mathrm{d.o.f}=1.02$ in both the 4- and 5- parameter cases, showing that both the MAP and median parametrizations are good estimates when calculating the  proton structure functions.

The good agreement with the HERA data is also illustrated in Fig.~\ref{fig:reducedcrosssection} where we show a comparison to the reduced cross section data in a few selected virtuality bins. The central lines are the average values obtained by computing the cross section using many posterior samples, and the uncertainty band corresponds to two standard deviation variation. To allow for asymmetric uncertainty estimates, in this work we always calculate separately the 2 standard deviation ($2\sigma$) uncertainty band above and below the mean value. We note that the reduced cross section is generically underestimated at low $Q^2$. However, we still get a good $\chi^2$ when the correlated systematic uncertainties in the HERA data are taken into account. If the statistical and systematical uncertainties in the HERA data are simply added in quadrature, we get a somewhat larger $\chi^2/\mathrm{d.o.f}=2.2$ in both the 4- and 5-parameter case. See also discussion in Appendix~\ref{appendix:correlations} for the effect of the correlated uncertainties on the posterior distribution. 

\begin{figure}
    \centering
    \includegraphics[width=\columnwidth]{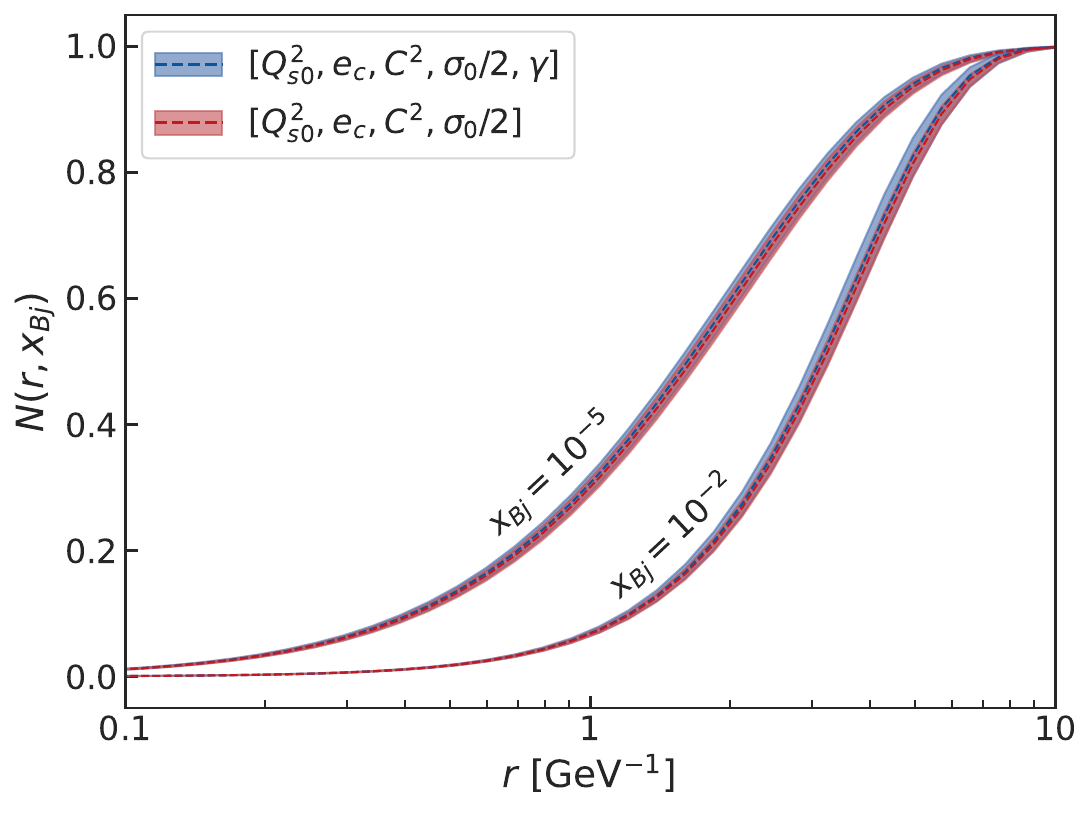}
    \caption{The dipole amplitude at the initial condition and evolved down to $\xbj=10^{-5}$ calculated in the 4- and 5- parameter cases. The band shows a $2\sigma$ uncertainty.}
    \label{fig:ieBK}
\end{figure}

The initial condition for the dipole amplitude $N(r,\xbj=0.01$) and the evolved amplitude at a much smaller $\xbj=10^{-5}$ are shown in Fig.~\ref{fig:ieBK}. Note that the initial dipole amplitude only depends on $\qso, e_c$ and $\gamma$, and furthermore the evolution speed is also sensitive to $C^2$. The central value of the dipole amplitude is obtained as an average of the dipole amplitudes computed using different samples from the posterior distribution, and the uncertainty estimate is a 2 standard deviation band. We show the dipole amplitudes both from the 4- and 5-parameter setups, and, as expected, the uncertainty band in the 5-parameter case is slightly larger due to a broader posterior distribution. However the actual (2 standard deviation) uncertainty in the dipole amplitude is  small, typically a few percent and maximally $\approx 10\%$.

\section{Applications}
 \label{sec:applications}

As the dipole-target scattering amplitude is a convenient degree of freedom at small-$x$ (see e.g. Ref.~\cite{Dominguez:2011wm}), the determined posterior distribution is a necessary input to calculations of all observables. The fact that the full posterior distribution, and not just the best fit values, are available, also enables us to rigorously take into account the uncertainties in the non-perturbative parameters describing the initial condition for the BK evolution. In this Section, we demonstrate how the uncertainties in the extracted dipole amplitude propagate to the inclusive quark-target cross section and to the nuclear modification factor for the $F_2$ structure function that will be measured at the EIC.

\subsection{Inclusive quark production}

Let us first consider quark-target scattering which is the relevant subprocess in inclusive forward hadron production in pp or pA collisions. 
The inclusive forward quark production cross section in proton-proton collisions is directly proportional to the two-dimensional Fourier transform of the initial dipole-amplitude~\cite{Dumitru:2002qt,Dumitru:2005gt,Albacete:2010bs,Lappi:2013zma}: 
\begin{equation}
    \frac{\dd\sigma^{q+A\to q+X}}{\dd{y}\dd[2]\kt} = \frac{\sigma_0/2}{(2\pi)^2} xq(x,\mu^2) \tilde S_p(\kt),
\end{equation}
where the produced quark transverse momentum and rapidity are $\kt$ and $y$, respectively. The dipole amplitude is evaluated at $x=|\kt|/\sqrt{s}e^{-y}$ where $\sqrt{s}$ is the center-of-mass energy. Furthermore, $xq(x,\mu^2)$ is the quark parton distribution function evaluated at scale $\mu^2$, and
\begin{equation}
    \tilde S_p(\kt) = \int \dd[2]{\rt} e^{-i\kt \cdot \rt} \left[ 1-N(\rt)\right].
\end{equation}

\begin{figure}
    \centering
    \includegraphics[width=\columnwidth]{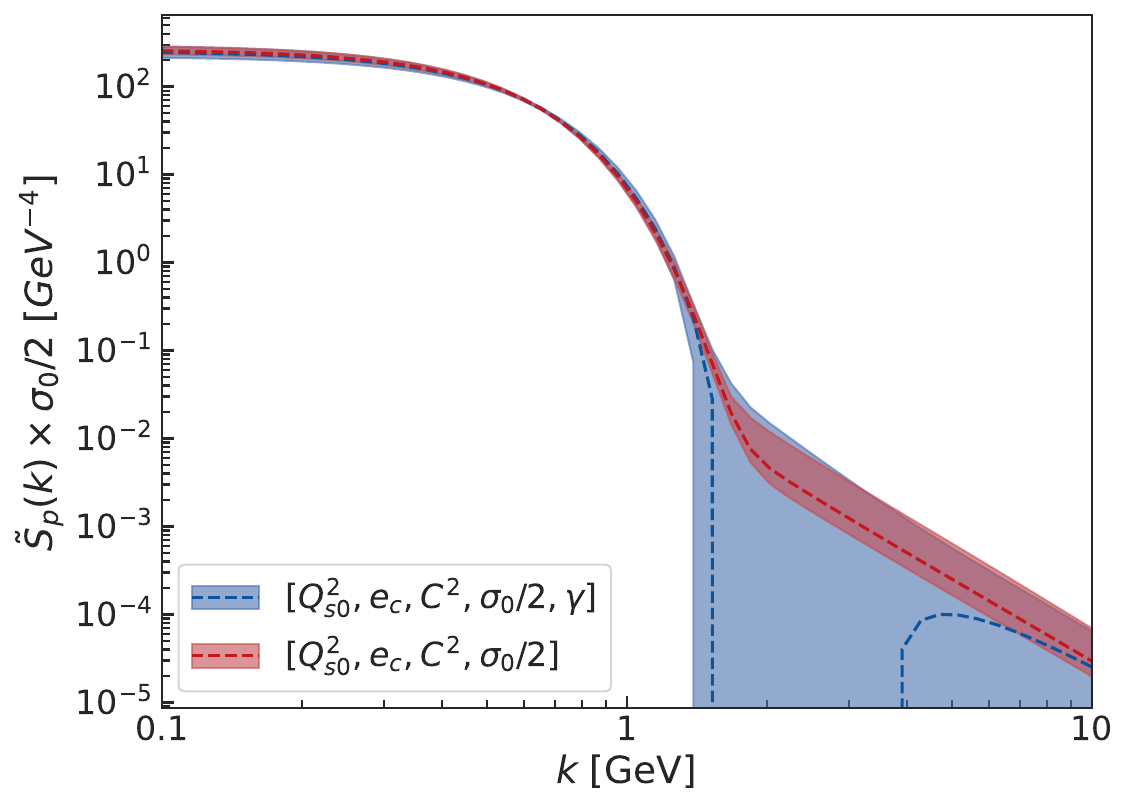}
    \caption{2D Fourier transform of the  dipole-proton amplitude scaled by $\sigma_0/2$ at the initial $x=0.01$ in the 4- and 5-parameter cases.} 
    \label{fig:2dft}
\end{figure}

The two-dimensional Fourier transform of the dipole amplitude scaled by $\sigma_0/2$ at the initial $\xbj=0.01$ is calculated in both the 4- and 5-parameter case, and the results are shown in Fig.~\ref{fig:2dft}. Although the median $\gamma$ obtained is close to 1, the posterior covers parametrizations where anomalous dimensions slightly larger than unity (up to $\gamma\approx 1.05$) are encountered, resulting in negative values of $\tilde{S_{p}}(\mathbf{k})$ in the large $\mathbf{k}$ region. This is because the Fourier transform is not positive definite with $\gamma>1$~\cite{Giraud:2016lgg}. At smaller $x$ after the BK evolution the Fourier transform would be positive definite as the BK evolution drives the anomalous dimension towards an asymptotic value $\gamma=0.6\dots 0.8$.

Additionally we observe that although the uncertainties in the determined dipole amplitude are typically around 5\% (see Fig.~\ref{fig:ieBK}), uncertainties in the quark-target cross section (in the $\tilde S_p \times \sigma_{0}/2$) can be significantly larger. Even in the 4-parameter case where the Fourier transform is positive definite, e.g. at $k=2\,\gev$ the upper limit of the $2 \sigma$ uncertainty band is around $150\%$ larger than the mean value. This highlights the importance of properly taking into account the parametrization uncertainties when applying the dipole amplitude determined from the HERA data when describing particle production processes at the LHC as e.g. in Ref.~\cite{Lappi:2013zma}. At next-to-leading order~\cite{Chirilli:2012jd,Stasto:2013cha} the possibility to emit a gluon changes the kinematics from $1\to 1$ at LO to $1\to 2$ at NLO, which can have a significant effect on the $p_T$ distribution (e.g. one can obtain a power-law tail at high $p_T$ even with a Gaussian dipole). The importance of properly take into account uncertainties in the initial dipole-target scattering amplitude when calculating inclusive spectra at NLO has been recently emphasize in Ref.~\cite{Mantysaari:2023vfh}.

\subsection{Nuclear effects on $F_2$}

Much more pronounced saturation effects can be expected to seen in scattering processes where the target is a heavy nuclei instead of the proton, as the saturation scale in the nucleus scales roughly as $\sim A^{1/3}$. 
The dipole-proton scattering amplitude determined in this work can be generalized to the dipole-nucleus case e.g. by employing the optical Glauber model. Following Ref.~\cite{Lappi:2013zma}, we write the initial condition for the BK evolution of the dipole-nucleus scattering amplitude at fixed impact parameter $\bt$ as  
\begin{multline}
    \label{eq:nuke-bk-ic}
    N_{A}(\rt, \bt, x = x_0) =
    1-\exp 
        \left[  - AT_{A}(\bt) \frac{\sigma_0}{2}
 \frac{\left(\rt^2\qso^2\right)^\gamma}{4} \right. \\
            \times \left. \ln \left( \frac{1}{|\rt| \Lambda_\text{QCD}}
             +  e_c\cdot  e   \right)   \right], 
\end{multline}
where $T_{A}(\bt)$ is the transverse thickness function of the nucleus of mass number $A$. The dipole-nucleus amplitudes at fixed $\bt$ are then evolved to small-$x$ by solving the BK equation without the impact parameter dependence. This thickness function is obtained by integrating the Woods-Saxon distribution 
\begin{equation}
    \rho_{A}(\bt, z) = \frac{n}{1 + \exp\left[ \frac{\sqrt{\bt^{2} + z^2} - R_{A}}{d} \right] }
\end{equation}
over $z$, where $d = 0.54$ fm and $R_{A} = (1.12A^{1/3} -0.86A^{-1/3})$ fm. The normalization condition $\int \dd[2]\bt T_A(\bt)=1$ fixes the overall constant $n$. 
In the large $|\bt| > b_\mathrm{max}$ region where $A T_A(\bt) \sigma_0/2 < 1$, the saturation scale of the nucleus falls below that of the proton. In order to avoid an unphysically rapid growth of the nuclear size in this region we write, following again Ref.~\cite{Lappi:2013zma}, $N_A(\rt,\bt,x) = A T_A(\bt) \frac{\sigma_0}{2} N(\rt,x)$, i.e. use a dipole-proton scattering amplitude scaled such that all nuclear effects vanish.

\begin{figure}
    \centering
    \includegraphics[width = \columnwidth]{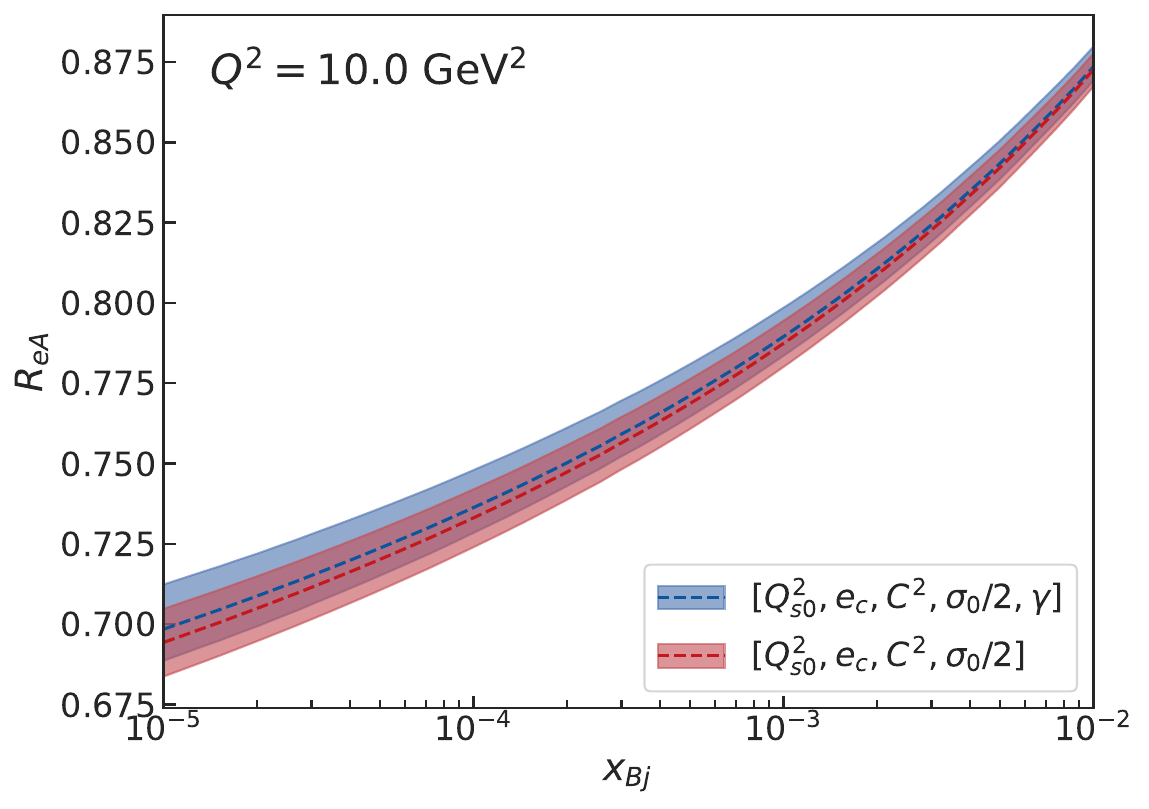}
    \caption{Nuclear modification factor for the structure function $F_2$ as a function of $x_{\mathrm{Bj}}$ at fixed $Q^2=10\,\mathrm{GeV}^2$.}
    \label{fig:rpa_xdep}
\end{figure}

\begin{figure}
    \centering
    \includegraphics[width = \columnwidth]{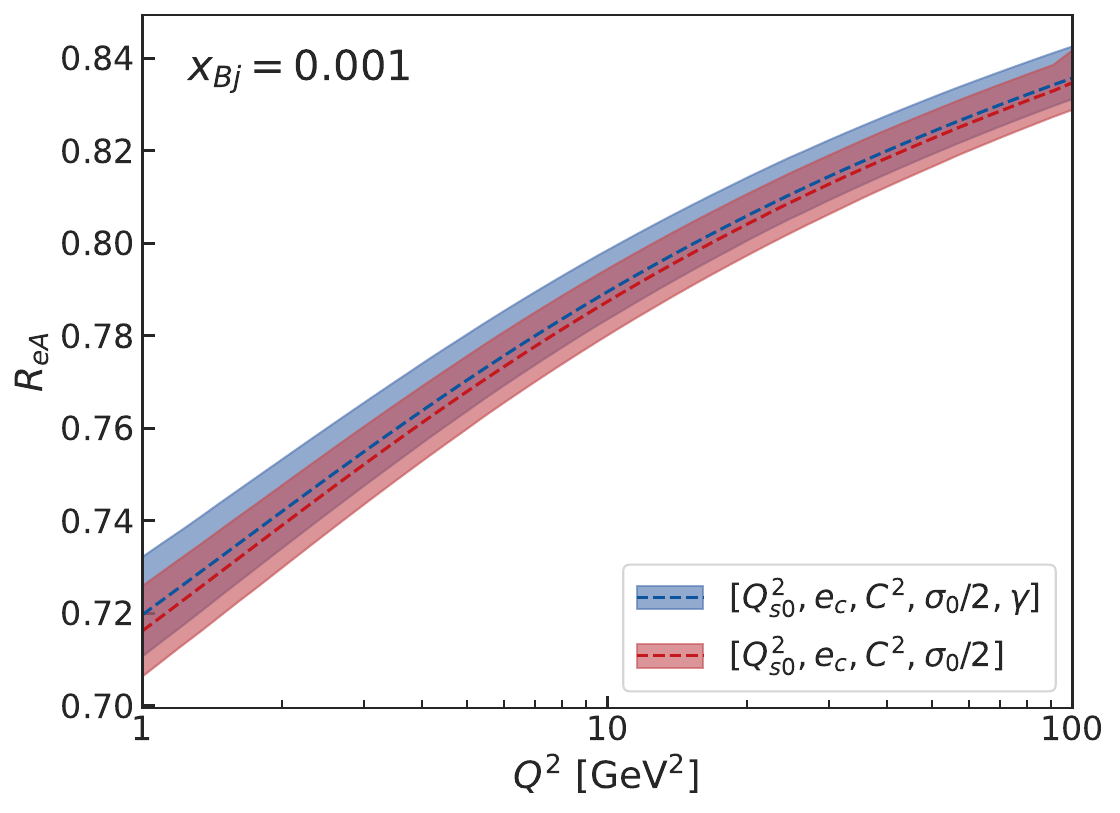}
    \caption{Nuclear modification factor for the structure function $F_2$ as a function of photon virtuality $Q^{2}$ at fixed $\xbj=0.001$. }
    \label{fig:rpa_q2dep}
\end{figure}

We compute the nuclear modification factor for the structure function $F_2$ defined as 

\begin{equation}
    R_{eA} = \frac{F_{2,A}}{A F_{2,p}},
\end{equation}
where $F_{2,A}$ is the $F_2$ structure function for the nucleus with mass number $A$. In this work we consider gold, i.e. $A=197$, for which this modification factor will be measured at the EIC.
The obtained $R_{eA}$ as a function of $\xbj$ is shown in Fig.~\ref{fig:rpa_xdep} and as a function of $Q^2$ in Fig.~\ref{fig:rpa_q2dep}. 

We predict a significant nuclear suppression in the $F_2$ structure function already in the EIC kinematics. By construction $R_{eA}\to 1$ at the initial $x=0.01$ in the limit $Q^2\to\infty$, and a stronger suppression is seen both towards small-$\xbj$ and  low-$Q^2$. Unlike above when studying the inclusive quark production cross section, in the case of $R_{eA}$ the parametrization uncertainties effectively cancel in the structure function ratio and the $2\sigma$ uncertainty is at 1--2\% level. The uncertainty grows slightly towards small-$\xbj$  where also the $C^2$ parameter controlling the evolution speed become important. 

\section{Conclusions and outlook}
\label{sec:conclusions}

We have determined the posterior distribution for the non-perturbative parameters describing the dipole-proton scattering amplitude at the initial condition of the small-$x$ BK evolution. This is a necessary input to  calculations of all scattering processes at high energy in the Color Glass Condensate framework. We have also determined, for the first time, uncertainty estimates for these non-perturbative parameters. The obtained posterior distribution makes it possible to rigorously take into account the uncertainties in the initial condition parametrization when looking for signals of gluon saturation from various different observables. The posterior distributions determined in this work can be found from the supplementary material.

We have shown that the HERA structure function data constrains most of the model parameters well and the remaining uncertainties in the obtained dipole amplitude are at a few percent level. The flexible Bayesian inference setup allows us to include more freedom in the initial dipole-proton amplitude than in the previous fits, and in particular we find that the HERA data does not require a large anomalous dimension $\gamma\approx 1.1\dots 1.2$ unlike in the analyses presented in Refs.~\cite{Lappi:2013zma,Albacete:2010sy,Albacete:2009fh}. Instead we get $\gamma \approx 1$ which is  more natural considering that $\gamma\le 1$ is required to obtain a positive definite quark production cross section and unintegrated gluon distribution~\cite{Giraud:2016lgg}. 

We demonstrate the uncertainty propagation by calculating the inclusive quark production cross section in proton-proton collisions and the nuclear modification factor $R_{eA}$ for the structure function $F_2$ to be measured at the EIC and at the LHeC/FCC-he. Significant uncertainties up to $\sim 100\%$ were found in the case of the inclusive quark production in proton-proton collisions. On the other hand, the uncertainties were found to mostly cancel in the nuclear modification factor $R_{eA}$ for the $F_2$ structure function. As such it can be crucial to properly propagate the uncertainties in the non-perturbative input when comparing the CGC predictions describing the gluon saturation effects to e.g. LHC inclusive spectra.  

In the future, we will apply the the developed computationally efficient setup to determine the initial condition for the BK evolution at next-to-leading order accuracy. This is especially intriguing as it has been recently shown that at NLO global analyses including both total and heavy quark production cross section data simultaneously become feasible~\cite{Beuf:2020dxl,Hanninen:2022gje}. We expect the proper treatment of initial condition uncertainties to be of equal importance as higher-order corrections when the Color Glass Condensate calculations are promoted to the precision level.

\begin{acknowledgments}
We thank J. Auvinen, T. Lappi and H. Paukkunen for useful discussions.
H.M and C.C. are supported by the Research Council of Finland, the Centre of Excellence in Quark Matter and projects 338263 and 346567. C.C. also acknowledges the support of the Vilho, Yrjö, and Kalle Väisälä Foundation. This work was also supported under the European Union’s Horizon 2020 research and innovation programme by the European Research Council (ERC, grant agreement No. ERC-2018-ADG-835105 YoctoLHC) and by the STRONG-2020 project (grant agreement No. 824093). 
Computing resources from the Finnish Grid and Cloud Infrastructure (persistent identifier \texttt{urn:nbn:fi:research-infras-2016072533}) were used in this work.
The content of this article does not reflect the official opinion of the European Union and responsibility for the information and views expressed therein lies entirely with the authors. 
\end{acknowledgments}

\bibliographystyle{JHEP-2modlong.bst}
\bibliography{refs}

\appendix 
\section{Treatment of Experimental Correlated Systematic Uncertainties}
\label{appendix:correlations}

In this work we have included, for the first time, the correlations between the different sources of systematical  uncertainty in the HERA structure function data when determining the initial condition for the BK evolution. In order to quantify the effect these previously neglected correlations  have on the determined initial condition, we also extract the initial condition for the BK evolution by neglecting all correlations in the experimental data. Instead, we choose the covariance matrix to be diagonal with  
$\Sigma_\mathrm{exp} = \mathrm{diag}(\sigma_1^{2}, \sigma_2^{2}, \sigma_3^{2}, ... , \sigma_N^{2})$ where the statistical and both correlated and uncorrelated systematical uncertainties are added in quadrature: 
 $\sigma_i^{2} = \sigma^{2}_\mathrm{sys,uncorr} + \sigma^{2}_\mathrm{stat,uncorr} + \sigma^{2}_\mathrm{corr}$.

\begin{figure*}[ht]
    \centering
    \includegraphics[width=\textwidth]{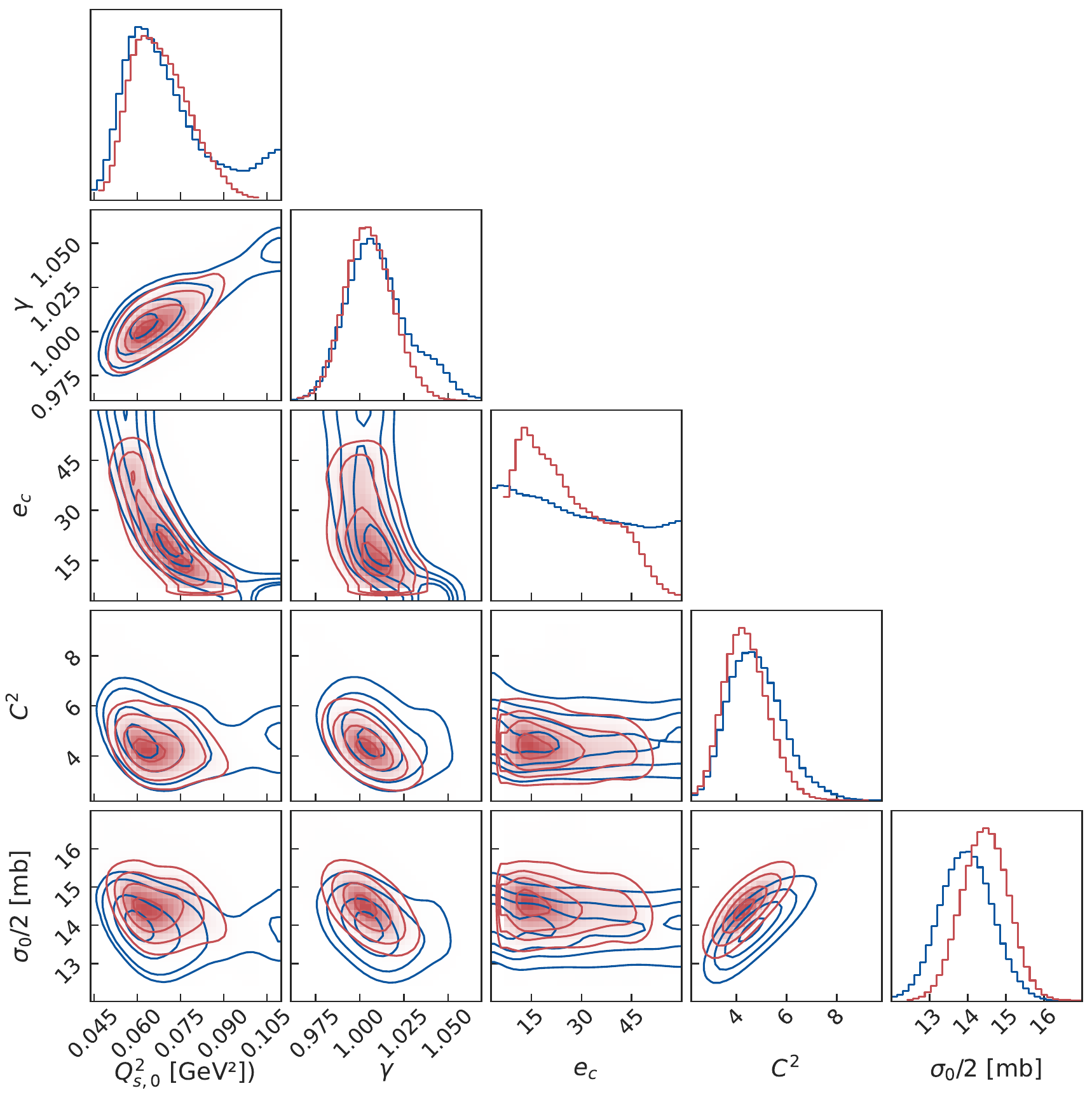}
    \caption{Posterior distribution in the 5-parameter case obtained with (blue) and without (red) including the correlations among the experimental uncertainties.} 
    \label{fig:corr_vs_uncorr}
\end{figure*}

 The posterior distribution obtained with and without the correlated uncertainties in the experimental data in the 5-parameter case are shown in Fig.~\ref{fig:corr_vs_uncorr}. Overall we find quite similar posterior distributions, except that  when the correlations between the experimental uncertainties are known there is more flexibility for the model calculations to agree with the data and we typically obtain broader distributions for the parameters. Additionally the preferred values for the $C^2$ (controlling the evolution speed) and $\sigma_0/2$ (proton size) are also slightly shifted by the inclusion of the experimental covariance matrix.

\end{document}